\documentclass[5p]{elsarticle}
\usepackage[utf8]{inputenc}
\usepackage[binary-units]{siunitx}
\usepackage{hyperref}
\usepackage{subcaption}
\usepackage{graphicx}
\usepackage{float}

\title{BDAQ53, a versatile pixel detector readout and test system for the ATLAS and CMS HL-LHC upgrades}

\author[1]{M.~Daas\corref{cor1}}\ead{daas@physik.uni-bonn.de}
\author[1]{Y.~Dieter}\ead{dieter@physik.uni-bonn.de}
\author[1]{J.~Dingfelder}\ead{dingfelder@physik.uni-bonn.de}
\author[1]{M.~Frohne}\ead{frohne@physik.uni-bonn.de}
\author[1]{G.~Giakoustidis}\ead{giakoustidis@physik.uni-bonn.de}
\author[1]{T.~Hemperek}\ead{hemperek@physik.uni-bonn.de}
\author[1]{F.~Hinterkeuser}\ead{hinterkeuser@physik.uni-bonn.de}
\author[1]{F.~H\"{u}gging}\ead{huegging@physik.uni-bonn.de}
\author[1]{J.~Janssen}\ead{janssen@physik.uni-bonn.de}
\author[1]{H.~Kr\"{u}ger}\ead{krueger@physik.uni-bonn.de}
\author[1]{D.-L.~Pohl}\ead{pohl@physik.uni-bonn.de}
\author[1]{P.~Rymaszewski}\ead{rymaszewski@physik.uni-bonn.de}
\author[1]{M.~Standke}\ead{standke@physik.uni-bonn.de}
\author[1]{T.~Wang}\ead{t.wang@physik.uni-bonn.de}
\author[1]{M.~Vogt}\ead{vogt@physik.uni-bonn.de}
\author[1]{N.~Wermes}\ead{wermes@uni-bonn.de}

\cortext[cor1]{Corresponding author}

\address[1]{Physikalisches Institut, Universit\"{a}t Bonn, Nussallee 12, 53115 Bonn, Germany}

\begin{document}
    \begin{abstract}
        BDAQ53 is a readout system and verification framework for hybrid pixel detector readout chips of the RD53 family. These chips are designed for the upgrade of the inner tracking detectors of the ATLAS and CMS experiments. BDAQ53 is used in applications where versatility and rapid customization are required, such as in laboratory testing environments, test beam campaigns, and permanent setups for quality control measurements.
        It consists of custom and commercial hardware, a Python-based software framework, and FPGA firmware. BDAQ53 is developed as open source software with both software and firmware being hosted in a public repository.
    \end{abstract}

    \date{\today}
    \maketitle

    \section{Introduction}
    Following the upgrade of the Large Hadron Collider~(LHC) to the High-Luminosity LHC (HL-LHC), many detector systems of the LHC experiments are upgraded to cope with the increased particle rates, fluences, and radiation dose. ATLAS~\cite{atlas} and CMS~\cite{cms} are two experimental installations at the LHC, built to record high-energy proton-proton collisions.
    Their particle trackers will be replaced by new all-silicon tracking detectors, which have been developed for several years and are currently being intensively tested and characterized~\cite{rd53a-results-analog, rd53a-results-powering}.
    The innermost parts of these tracking systems, the pixel detectors, employ hybrid pixel detector technology where the sensing element and the readout chip 
are separate ASICs. The readout chip in particular faces a remarkable task given the high-rate and high-radiation environment of the HL-LHC.

Therefore, the readout chip for the pixel detectors of both ATLAS and CMS is developed in a collaborative effort by the RD53 Collaboration at CERN~\cite{rd53}.
    The first large-scale prototype chip produced by this collaboration is called RD53A~\cite{rd53a}, while the first candidates for the production chips for the upgraded ATLAS and CMS detectors are named ITkPix-V1 and CROC\_V1, respectively.
    
    These novel readout chips come with many new features, like higher readout bandwidth and new data formats which demand new readout systems to interface, test and characterize them and the assembled hybrid pixel modules. This not only concerns electronic tests of the ASICs, but also comprises intensive characterizations of full-size pixel detector modules during laboratory tests with radioactive sources as well as dedicated test beam campaigns.

The RD53-based pixel detector modules of ATLAS and CMS can likely be regarded as very prominent pixel detector systems for years to come, with use cases beyond their applications at the LHC. Therefore the readout and test system presented in this paper represents an important backbone also for further  developments of RD53-based hybrid pixel detector systems beyond their primary usage.       
    
    \section{BDAQ53 Readout System}
    BDAQ53\footnote{BDAQ53 on CERN Gitlab:\newline \url{https://gitlab.cern.ch/silab/bdaq53}}~\cite{bdaq} is a versatile readout system and verification framework for the family of readout chips designed by the RD53 collaboration.
    
    Since this new generation of readout chips uses a new command interface and is capable of data transfer at \SI{5}{Gbit/s}, more than ten times faster than its predecessors~\cite{usbpix}, a new readout system has been developed to interface with these chips.
    
    BDAQ53 constitutes the basis for communication with the readout chip. Hardware interfaces are provided by the custom-designed, FPGA-based hardware platform of BDAQ53, while the Python-based software framework running on a connected PC enables granular control, calibration and data taking.
    
    Common use cases of BDAQ53 include tabletop lab measurements like chip or sensor characterization, test beam campaigns, as well as stationary setups for example for wafer probing and module production tests. BDAQ53 aims to stay as lightweight and versatile as possible.
    
    \begin{table*}[h]
    	\centering
    	\begin{tabular}{lll}
    		\hline
    		& Requirements&BDAQ53 Specification \\
    		\hline
    		\# of Data Lanes & 1 (up to 4) per chip & 7 \\
    		Data Lane Bandw. & up to \SI[per-mode=symbol]{1.28}{\giga\bit\per\second} & \SI[per-mode=symbol]{640}{\mega\bit\per\second}, \SI[per-mode=symbol]{1.28}{\giga\bit\per\second} \\
    		Jitter of gen. \texttt{CMD} & - & $\text{TIE} = \SI{16}{\pico\second} \text{ (RMS)}$ \newline
    		at $\text{BER} = \num{e-12}$ \\
    		\hline
    		Optional Features & \multicolumn{2}{l}{Data Buffer, CDR Bypass Mode, Multi-Chip Readout} \\

    	\end{tabular}
    	\caption{Requirements of RD53 readout chips and specifications of BDAQ53}
    \end{table*}
    
    \section{Hardware}
    
    \begin{figure}[htb]
        \centering
        \includegraphics[width=0.48\textwidth]{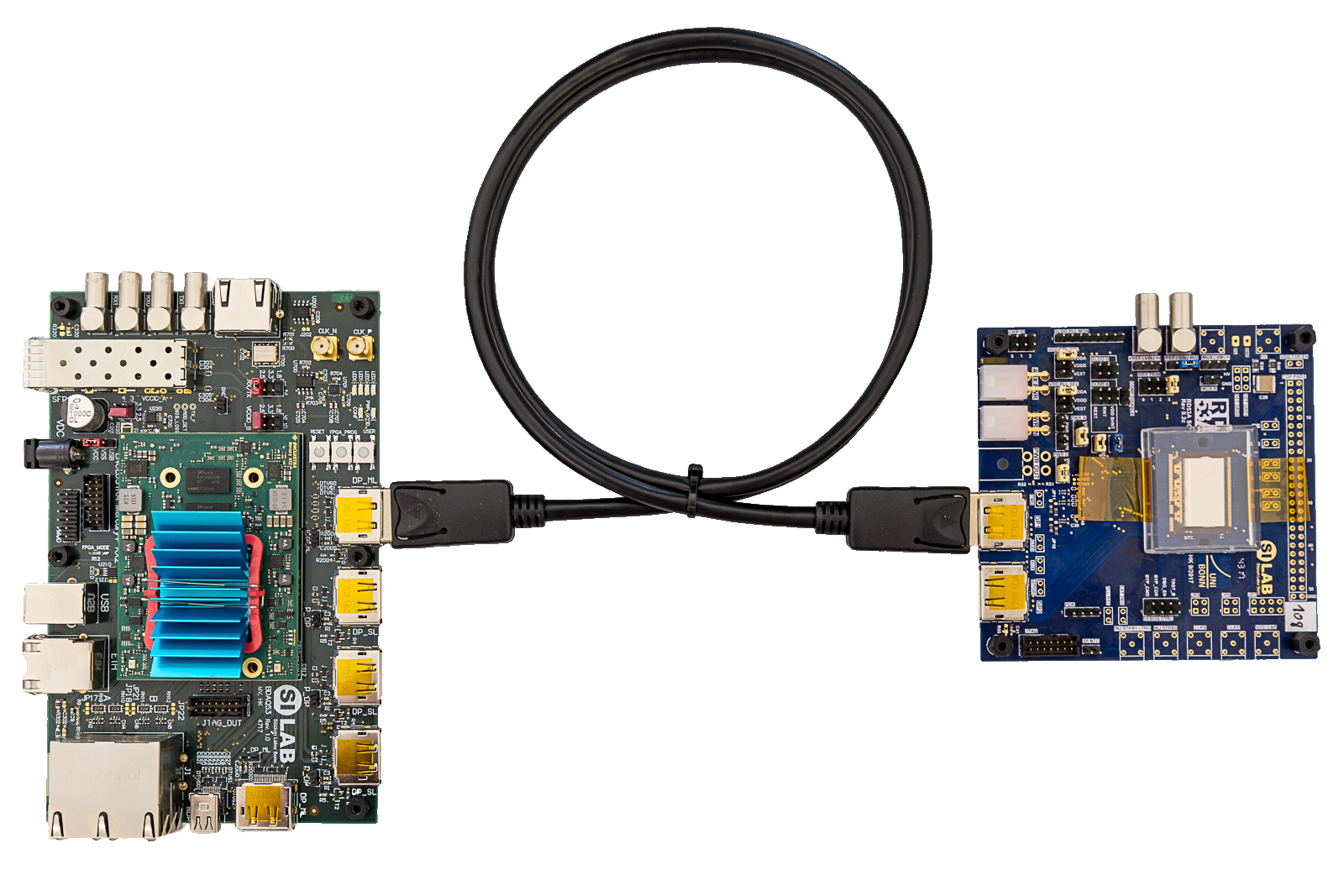}
        \caption{BDAQ53 base board (left) with Mercury+ KX2 daughter board, connected to an RD53A Single Chip Card (right) via DisplayPort.}
        \label{fig:bdaq}
    \end{figure}
    
    Figure~\ref{fig:bdaq} shows the custom base board that was designed to accommodate a commercially available FPGA daughter board and provide hardware interfaces to the device under test (DUT), the readout PC, and optional periphery. The base board also includes programmable clock generator chips, signal level translators and a multiplexed analog front-end for temperature measurements.
    The 4-layer PCB features impedance controlled differential lines and ESD-protection diodes close to the DisplayPort connectors. The DisplayPort standard was chosen by the RD53 collaboration for its high rate capability, high availability and low price.
    
    The Mercury+ KX2~\cite{enclustra} module was chosen as FPGA daughter board for BDAQ53, since it is a commercial product with long-term availability. It houses a Xilinx Kintex7~\cite{kintex7} FPGA, whose resources are utilized to about \SI{80}{\percent} and additionally contains several voltage regulators, which provide enough current to power the FPGA as well as the active base board components of BDAQ53.
    The KX2 module provides access to 8 high speed transceiver channels of the FPGA, which are used for communication with the DUT via the Xilinx Aurora protocol~\cite{aurora}, the implemented data encoding of RD53A and its successors. Four receiver channels are grouped into one multi-lane DisplayPort, while 3 single-lane ports are each routed to a single receiver. The remaining transceiver is connected to an SFP+ slot on the base board and can be used as a high-speed data interface to the DAQ PC, which supports data rates of up to \SI{10}{Gbit/s}.
    
    Apart from the custom-designed PCB, BDAQ53 also supports other hardware platforms, for example the commercially available Xilinx KC705 development board.

    \subsection*{Firmware}
\begin{figure}[htb]
        \centering
        \includegraphics[width=0.48\textwidth]{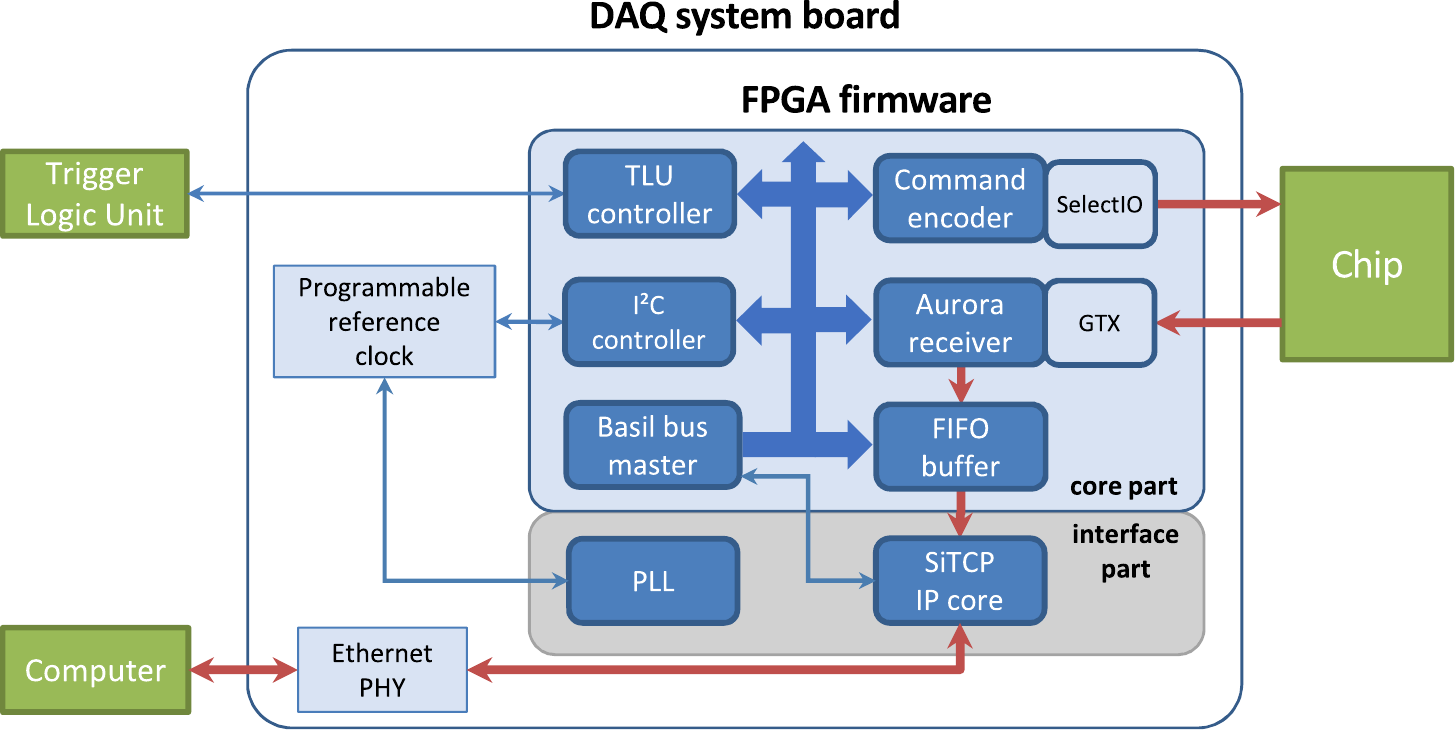}
        \caption{Functional firmware and DAQ hardware blocks. Main data flow depicted with red and control flow with blue arrows, respectively.}
        \label{fig:fig_daq}
    \end{figure}

    The firmware for BDAQ53 is written in Verilog and a functional block diagram is shown in Fig.~\ref{fig:fig_daq}. Due to the separation of the core firmware components and the input/output blocks, the core firmware can be integrated into a simulation testbench, as described in section \ref{sec:testbench}. The core firmware contains the functional elements necessary for basic operation. It does not include the main system clock PLL and the Ethernet interface, which are handled by the respective I/O module.
    This modularity also simplifies portability to different FPGA types or readout boards.

    Common firmware modules like FIFO buffers, a trigger- and a Time to Digital Converter (TDC) module are instantiated from the basil framework~\cite{basil}, while the RD53-specific command encoder and the Aurora receiver are part of BDAQ53. The firmware modules are connected to a shared internal control bus, which is also provided by basil.
    
    Current versions of the BDAQ53 firmware utilize about \SI{25}{\percent} of the FPGAs Look-Up Tables (LUTs) and almost \SI{80}{\percent} of the available Block Random Access Memory (BRAM), which is used as fast buffer memory for the chips data lanes and for the outgoing Ethernet connection.

    A script controls the firmware build process, since every base board and configuration variant requires a different set of parameters. A list covering the supported combinations is included in the script. During the synthesis process, these parameters determine if certain firmware features are included in the resulting firmware or how many entities of specific blocks are instantiated.

    \subsection*{Interface to Software}
    Two independent communication channels are used to transfer data between the readout board and the PC, sharing the same Ethernet interface and IP address, but using different protocols. User Datagram Protocol (UDP) is used to access the basil bus and configure and control the firmware modules, to benefit from its low overhead and latency.
    Data words from the Aurora receivers and additional modules are tagged with a data type header, before being merged into a common FIFO buffer and sent to the DAQ PC via Transmission Control Protocol (TCP). Each Aurora word is tagged with a channel ID to identify the data source, i.e., the DUT connected to the particular receiver.
    This architecture enables full control over raw data words, interpretation and event building in software, which is necessary to evaluate the data format and detect transmission errors.

    \section{Software Framework}
	\begin{figure}[htb]
        \centering
        \includegraphics[width=0.48\textwidth]{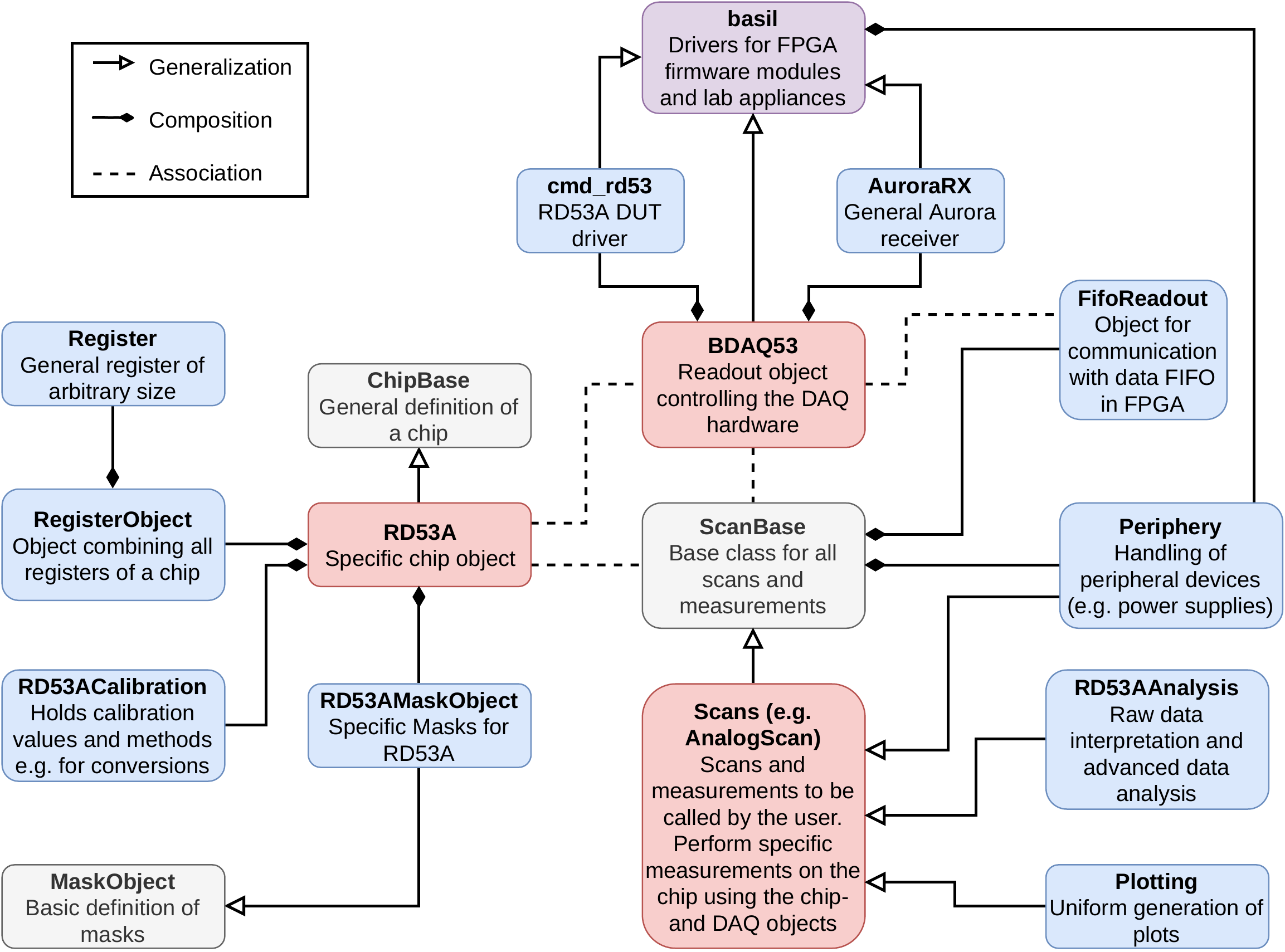}
        \caption{Class diagram of BDAQ53. The most important classes are shown in red, base classes are shown in white. The basil framework shown in purple provides several generic base classes for hardware interfaces.}
        \label{fig:bdaq_uml}
    \end{figure}    
    
    The Python-based software framework of BDAQ53 is developed with modularity and rapid development in mind. Figure~\ref{fig:bdaq_uml} shows the overall software scheme, which illustrates how the different classes interact with each other. The software consists of a chip class that is specific to the chip of the DUT, for example RD53A. This class handles all the communication with the chip, chip-specific commands, default configuration settings and calibration constants.
    
    A separate class for the readout system, called \textit{bdaq53} handles functions and commands that are specific to the readout hardware. It also accommodates functions which might be specific to the used hardware platform. Since chip communication is routed via the readout system hardware, the chip class accesses a subset of methods of the readout system class.
    
    The last essential part of the software is the scan or measurement scripts. These scripts exclusively contain the algorithm of a specific scan which is independent of the chip type and hardware platform. Scan scripts inherit most methods from a base class called \textit{scan\_base}, which incorporates all necessary steps to perform a scan on a chip. These basic steps are structured into an \textit{initialization} and \textit{configure} step, which ensure that communication is established and the chip is configured correctly. After these first steps, the algorithm defined in the scan script is executed and in the end, an optional \textit{analyze} step can be implemented to process and visualize the data, as shown in the following section.
    
    In addition, more classes are implemented for handling of specific optional tasks. For example, the \textit{firmware\_manager} class can be used to automatically download and flash the latest firmware to the FPGA before a scan is executed, which can be very useful for automated tasks. Another additional class is called \textit{periphery} and is used to handle and control peripheral devices like power supplies. This way it can be ensured that the chip is correctly powered on, or even power cycled before a scan starts. Both features are necessary for unattended and automated testing, which is a requirement for testing large batches of chips for tasks like quality control during mass production.
    
    \subsection*{Data Handling and Analysis}
    Once the readout thread in the software is started, the raw data words received from the FPGA via the Ethernet interface are written to disk in HDF5 format, a data format for efficient storage of large amounts of scientific data~\cite{hdf5}. This allows the meta data of a scan, for example a readout block ID, timestamps, chip configuration and more, to be stored together with the raw data in a single file. In normal operation, all interpretation of the data is handled offline by an \textit{analysis} class, after the scan is finished. The analysis is executed by the scan script subsequent to the scan, but can also be invoked manually any time after the scan without the need for any hardware being connected to the PC, as long as the raw data file is available. A class for online analysis of data is optionally available, which can be used in tuning routines to react to results of a previous iteration, for example for threshold tuning algorithms based on binary search. The offline analysis routine creates another HDF5 file for the interpreted data including recorded hits with pixel coordinates, timestamp and Time-over-Threshold (ToT) values. Furthermore, different histograms are already created and stored in the interpreted-data file, to allow for fast and easy graphical representation of the data.
    
    In addition, every scan outputs a configuration file and, if applicable, a mask file, which can be used to rerun the scan with the exact same settings as before.
    
    \subsection*{Data Visualization} \label{sec:visualization}
    In the analysis process, a PDF file is created that contains a relevant set of plots depending on the type of scan. The first page of this document summarizes the type of scan that was performed, the chip's serial number, the time of execution, all analog chip and DAQ settings necessary to reproduce the scan, as well as the software version of BDAQ53 that was used to generate the data file. Furthermore, a histogram is created showing a categorized number of errors, which can be helpful for DAQ or chip fault diagnostics. In addition, a set of relevant plots visualizing the data of the scan is produced. These can include timing related plots such as the distribution of relative hit timings, threshold or hit values of individual pixels, which are shown as an x-y hitmap with the information in question coded via the color scale, as well as two- or three-dimensional histograms of many other variables.
    
    \begin{figure*}[htb]
        \centering
        \begin{subfigure}[t]{0.45\textwidth}
            \centering
            \includegraphics[width=\textwidth]{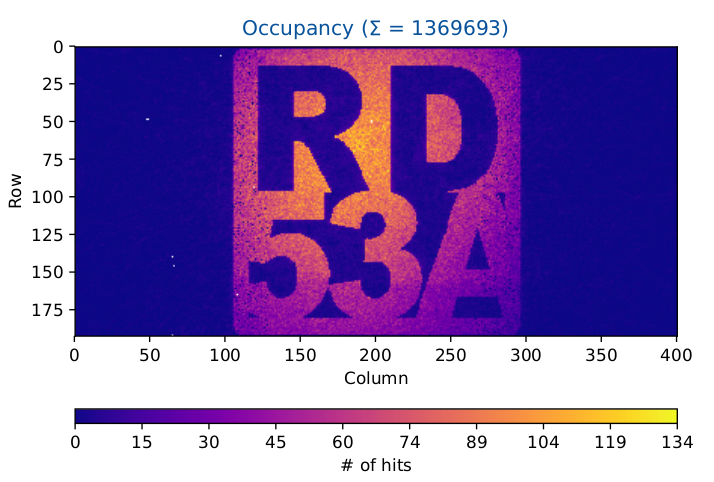}
            \caption{Hitmap of a BDAQ53 \textit{self-trigger scan} in a particle beam. The image was generated by running an RD53A assembly in self-trigger mode, accepting only triggers in the shape of the RD53A logo.}
            \label{fig:logoscan}
        \end{subfigure}
        \quad
        \begin{subfigure}[t]{0.45\textwidth}
            \centering
            \includegraphics[width=\textwidth]{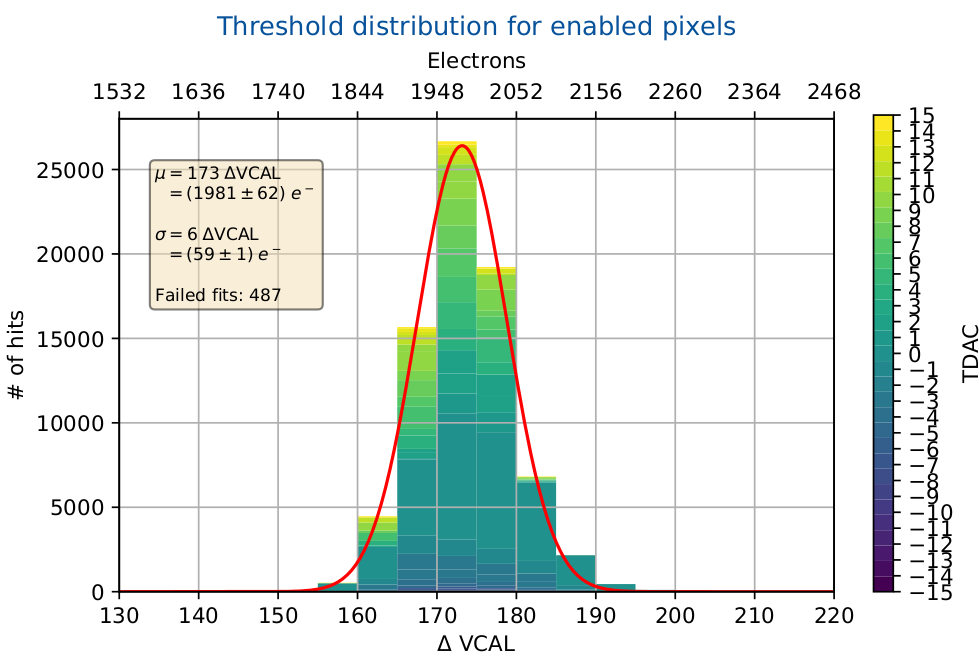}
            \caption{Threshold distribution after tuning a full RD53A chip to a threshold of \SI{2000}{e^-}. The color scale shows the TDAC distribution within the individual bins.}
            \label{fig:threshold}
        \end{subfigure}
        \caption{Example plots from different BDAQ53 scans of a RD53A single chip assembly.}
        \label{fig:example_plots}
    \end{figure*}
    
    Figure \ref{fig:example_plots} shows two exemplary plots created by BDAQ53. Figure~\ref{fig:logoscan} shows the result of a \textit{self-trigger scan} performed during a test beam campaign at the CERN SPS\footnote{Super Proton Synchrotron}. Triggers are only accepted by specific pixels defined by the RD53A logo. Hits in these pixels generate triggers, that initiate a readout of all hit pixels. This results in the actual beam profile being shown in the shape of the logo that is used for masking other pixels. Figure \ref{fig:threshold} shows one of the plots generated by a \textit{threshold scan} that was performed after tuning all three analog front-ends of an RD53A chip individually to a threshold of \SI{2000}{e^-}. It shows a histogram of the local threshold distribution of all enabled pixels measured by injections generated by the chip. Additionally, the color scale shows the distribution of local threshold DAC values within each bin of the histogram.

    \section{Simulation Environment} \label{sec:testbench}
    Test-driven development is a widely used process in software development to improve productivity, as it shifts the effort of trouble-shooting to the development of reusable tests.
    
	\begin{figure}[htb]
        \centering
        \includegraphics[width=0.48\textwidth]{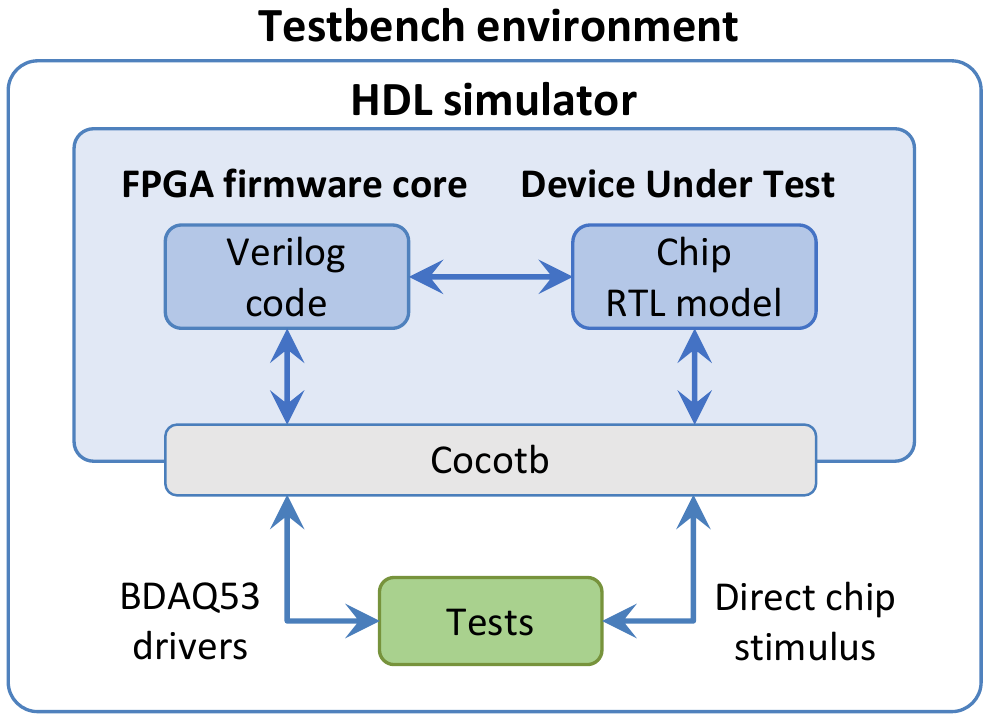}
        \caption{Simulation environment, based on an HDL simulator and a Cocotb interface layer to connect to test routines written in Python.}
        \label{fig:fig_sim}
    \end{figure}    
    
    This method is easy to implement for projects written in a single language, but requires additional effort in more complex scenarios. For example, a realistic BDAQ53 use case scenario consists at least of the Python software, the Verilog FPGA firmware and the System Verilog models of the RD53 chip's digital logic. These generally incompatible languages are operated together by using a co-simulation framework like Cocotb~\cite{cocotb}. Cocotb acts as an interface between the tests written in Python and a Verilog simulator which runs the firmware and chip model, replacing the physical Ethernet interface and chip of an actual setup. A schematic overview of this environment is shown in Fig.~\ref{fig:fig_sim}.
    
    The main benefit of this technique is the ability to create a realistic simulation scenario, which can be used to verify both the readout system and the chip model. Additionally, it enables DAQ development based on simulation, before the chip is physically available. For RD53A, this meant that basic configuration routines and injection tests could be developed and tested already during the chip development and production phase.
    Issues in the RD53A digital design related to the Aurora communication could be identified and fixed prior to chip submission and the DAQ system was able to take first data from RD53A only a few hours after the first sample was available.
    
    As part of the continuous integration, automated functionality tests of all software modules are integrated into the BDAQ53 software deployment process. These tests are started automatically after every change or addition to the repository and passing certain tests is mandatory for merging new code into the main branches. Apart from purely software-based tests evaluating for example the interpretation of given raw data files or the integrity of generated commands, all scans and modules are also tested against a full simulation of the chip's digital logic. More test suites cover running the software on a dedicated machine with a connected chip, as well as the integration with other frameworks such as EUDAQ~\cite{eudaq}. This procedure ensures high availability and operational readiness.
       
    \section{Particular Features}
    The following paragraphs highlight some distinctive features of BDAQ53 that qualify the readout system for use in chip or sensor characterization tasks, quality assurance and control, and operation in lab or test beam environments.
    
	\subsection*{HitOr Triggering}
    Self-triggered operation is a simple way to characterize assemblies with external particle sources without the need for an external triggering mechanism. Since RD53A does not provide this functionality on-chip, BDAQ53 is able to generate triggers from the chip's \mbox{HitOr}, a logical OR of all pixels' discriminator outputs, thus enabling effective self-triggered operation of RD53A-based assemblies. This functionality is achieved by means of a trigger state machine implemented in the FPGA that generates triggers based on the pulses on the HitOr line which are sent directly to the chip with the correct latency and a configurable vetoing mechanism.
    
    \subsection*{Multi-Chip Readout}
    Multi-Chip readout refers to the ability to connect up to four chips to a single BDAQ53 board at the same time. With an additional multiplexer card, up to four quad-chip assemblies, consisting of 4 readout chips each, can be connected to BDAQ53 for automatic testing. In case of external or self-trigger scans, all four chips are read out in parallel to enable for instance measurements with radioactive sources or in test beams with substantial time savings.
    
    \subsection*{TDC Method}
    With the TDC method~\cite{tdc}, the width of the pulses on the \mbox{HitOr} line of the DUT is sampled by the FPGA of the readout system with a \SI{640}{\mega\hertz} clock. This allows for much finer sampling of the hit pulses than is possible using the built-in 4-bit ToT mechanism, providing a higher resolution charge and hence energy measurement.
    High-resolution energy measurements are essential for charge calibration and profound characterization of passive sensors using an RD53 readout chip.
    Information achieved with this method was used to characterize and calibrate the charge injection circuit of the RD53A readout chip.
    
    \section{Use Cases}
    Typical use cases for BDAQ53 include test beam campaigns, where beam time is scarce and expensive and as little time as possible should be dedicated to setting up and configuring the readout system. On the other hand, versatility and easy integration of peripheral devices are features which help with the development of stationary setups, for example for testing readout chips on wafer level on a probe station or for defined quality control measurements.
    
    \subsection*{Test Beams}
    Several successful test beam campaigns have been conducted using BDAQ53 at multiple facilities including the CERN SPS and DESY. A first campaign took place in May 2018, shortly after first assemblies based on RD53A were available. This campaign was dedicated to testing and optimizing all features necessary for operating RD53A in beam together with a beam telescope such as ACONITE~\cite{aconite}. BDAQ53 is fully compatible and regularly used with the EUDAQ DAQ framework~\cite{eudaq}, supporting all three handshake methods.
    
    Multiple test beam campaigns in the light of the ATLAS and CMS HL-LHC upgrades have been conducted successfully using BDAQ53, enabling valuable characterization results of different sensor prototypes~\cite{cmos, 3d}.
    
    \subsection*{Wafer Probing}
    The first step in module mass production for upcoming detector upgrades is chip testing on wafer level. To enable these chip tests, a wafer probing setup has been developed at the University of Bonn, using basil to control peripheral devices such as power supplies and the probe station itself, as well as BDAQ53 for communication with and testing of the RD53A chips. This setup was also duplicated and is now used at multiple sites for distributed mass testing. Between the different testing sites, more than 7000 RD53A chips (83 wafers) have been tested in total using this setup.
    
    \subsection*{Module Quality Control}
    In preparation for mass production of modules for the upgrade of the ATLAS Inner Tracker, a test setup based on BDAQ53 is being developed and will be used for mass testing modules after assembly. In this important step for quality control, the performance of assembled modules is verified by following a specific testing routine under well-defined environmental conditions.
    
    \section{Conclusion}
    BDAQ53 is a versatile and lightweight readout system for RD53-like front-end chips for hybrid silicon pixel detectors. It has matured over the course of two years characterizing and evaluating RD53A, the first large scale prototype readout chip of the RD53 collaboration. Future variants of this chip, like the ATLAS ITkPix-V1 and CMS CROC\_V1 will be supported as well and the simulation environment of BDAQ53 is already used to aid in chip design.
    
    The system is based on a commercial FPGA daughter board on a custom PCB or fully commercial PCBs and includes Verilog firmware and a Python-based software framework. It features simultaneous readout of multiple chips, self-triggering for RD53A and advanced features like charge measurements with increased resolution based on the TDC method.
    
    It has been successfully used in multiple test beam campaigns at different facilities and is continuously used in stationary setups for testing and quality control~\cite{3d, usage:monteil, usage:padras, usage:beyer, usage:vogt19, usage:duarte19}.
    
    \section{Acknowledgments}
    This project has received funding from the German Ministerium
f\"{u}r Bildung, Wissenschaft, Forschung und Technologie (BMBF) under contract
{no. 05H15PDCA9} and the H2020 project AIDA-2020, under grant agreement {no. 283~654168}.

\end{document}